# Observation of shell effects in superconducting nanoparticles of Sn

Sangita Bose[1*], Antonio M. García- García[2,3*], Miguel M. Ugeda[1,4], Juan D. Urbina[5], Christian H. Michaelis[1], Ivan Brihuega[1,4*] and Klaus Kern[1,6]

[1] Nanoscale Science Department, Max Planck Institute for Solid State Research, Heisenbergstrasse 1, Stuttgart, D-70569, Germany.

[2] Physics Department, Princeton University, Princeton, New Jersey 08544, USA.

[3] CFIF, Instituto Superior Técnico, UTL, Av. Rovisco Pais, 1049-001 Lisboa, Portugal.

[4] Univ. Autonoma Madrid, Dept. Fis. Mat. Condensada, E-28049 Madrid, Spain.

[5] Institut für Theoretische Physik, Universität Regensburg, D-93040 Regensburg, Germany.

[6] Institut de Physique de la Matière Condensée, Ecole Polytechnique Fédérale de Lausanne, CH-1015 Lausanne, Switzerland.

*In a zero dimensional superconductor quantum size effects (QSE)[1,2] not only set the limit to superconductivity but are also at the heart of novel phenomena like shell effects, which have been predicted to result in large enhancements of the superconducting energy gap[3,4,5,6]. Here, we experimentally demonstrate these QSE through measurements on single, isolated Pb and Sn nanoparticles. In both systems superconductivity is ultimately quenched at sizes governed by the dominance of the quantum fluctuations of the order parameter. However, before the destruction of superconductivity, in Sn nanoparticles we observe giant oscillations in the superconducting energy gap with particle size leading to enhancements as large as*

[*] To whom correspondence should be addressed: sangita.bose@fkf.mpg.de, ag3@Princeton.EDU ivan.brihuega@uam.es

*60%. These oscillations are the first experimental proof of coherent shell effects in nanoscale superconductors. Contrarily, we observe no such oscillations in the gap for Pb nanoparticles, which is ascribed to the suppression of shell effects for shorter coherence lengths. Our study paves the way to exploit QSE in boosting superconductivity in low dimensional systems.*

Downscaling a superconductor and enhancing superconductivity has been a major challenge in the field of nanoscale superconductivity. The advent of new tools of nanotechnology for both synthesis and measurement of single, isolated mesoscopic superconducting structures has opened up the possibility to explore novel and fascinating phenomena at reduced dimensions[7,8,9,10,11,12,13,14,15,16] One of them, the parity effects in the superconducting energy gap, was demonstrated almost two decades ago in the only experiments which have been able to access the superconducting properties of an individual nanoparticle[7] till date. Another exciting prediction is the occurrence of shell effects in clean, superconducting nanoparticles[4,5,6].

The origin of shell effects is primarily due to the discretization of the energy levels in small particles which leads to substantial deviations of the superconducting energy gap from the bulk limit. For small particles, the number of discrete energy levels within a small energy window (pairing region) around the Fermi energy ($E_F$) fluctuates with very small changes in the system size. Consequently this leads to fluctuations in the spectral density around $E_F$. Since in weakly coupled superconductors electronic pairing mainly occurs in a window of size $E_D$ (Debye energy) around $E_F$, an increase (decrease) of the spectral density around $E_F$ will make pairing more (less) favorable, thereby increasing (decreasing) the energy gap ($\Delta$). As a consequence the gap becomes dependent on the size and the shape of the particle (see schematic drawing in Fig. 1). The strength of fluctuations also increases with the symmetry of the particle, since symmetry introduces degeneracies in the energy spectrum. It is easy to see that these degenerate levels will enhance the fluctuations in the spectral density and also in the gap as the number of levels within $\pm E_D$ of $E_F$, and consequently the number of electrons taking part in paring, fluctuates dramatically. These degenerate levels will be referred to as 'shells' in analogy with the electronic and nucleonic levels forming shells in atomic, cluster and nuclear physics (see Ref. 3 and references therein). For cubic or spherical particles this might lead to a large modification of $\Delta$. Theoretically, these shell effects are described quantitatively by introducing finite size corrections to the BCS model[5,6]. In this letter, through our scanning tunneling spectroscopic measurements on individual superconducting nanoparticles of Pb and Sn, we demonstrate for the first time the existence of these shell effects and the influence of the superconducting coherence length on them.

Fig. 2**a** shows the schematic of the experimental measurement where an STM tip is used to measure the tunneling density of states (DOS) of superconducting nanoparticles of both Pb and Sn. A typical representative STM topographic image for Sn nanoparticles (for Pb nanoparticle topographic image, see Ref. 17 ) with varying size on a BN/Rh(111) substrate (see Methods for details) is shown in Fig. 2**b**. We take the height of the nanoparticle as our reference since it is measured with a high degree of accuracy with the STM. The quasiparticle excitation spectra (conductance plots of dI/dV *vs* V normalized at

+5mV) for a selection of Pb and Sn nanoparticles at a temperature of 1.2-1.4 K are plotted in Figs. 2**c-e**. We fitted each spectrum with the tunneling equation,[18]

$$G(V) = \left.\frac{dI}{dV}\right|_V = G_{nn} \frac{d}{dV}\left\{\int_{-\infty}^{\infty} N_s(E)\{f(E) - f(E - eV)\}dE\right\} \qquad (1)$$

Where $N_s(E)$ is the DOS of the superconductor, $f(E)$ is the Fermi-Dirac distribution function and $G_{nn}$ is the conductance of the tunnel junction for $V \gg \Delta/e$. $N_s(E)$ is given by:

$$N_s(E,\Gamma,T) = \mathrm{Re}\left[\frac{|E| + i\Gamma(T)}{\sqrt{(|E| + i\Gamma(T))^2 - \Delta(T)^2}}\right] \qquad (2)$$

Where, $\Delta(T)$ is the superconducting energy gap and $\Gamma(T)$ is a phenomenological broadening parameter which incorporates all broadening arising from any non-thermal sources (conventionally it is associated with the finite lifetime ($\tau$) of the quasiparticles, $\Gamma \sim \hbar/\tau$)[19]. There is an excellent agreement between the experimental data and the theoretical fits, giving unique values of $\Delta$ and $\Gamma$ (plotted as a function of particle size in Fig. 2**f**, **g** respectively). Comparing the raw data for the Pb and Sn, we observe that there is a gradual decrease in the zero bias conductance dip with particle size for Pb nanoparticles (Fig. 2**c**), while for Sn nanoparticles (Figs. 2**d**, **e**) there is a non monotonic behavior which strongly depends on the particle size regime. We observe that though the large Sn particles (>20 nm) differing in a size of 1 nm have similar DOS signifying similar gaps, there is a large difference in the DOS and hence $\Delta$, for the smaller Sn particles (< 15 nm) even if they differ by less than 1 nm in size. The difference in the two systems is brought out more clearly in Fig. 2**f** where we plot the normalized gap (normalized with respect to their bulk values). For Pb, $\Delta$ decreases monotonically with decrease in particle size while there is a huge variation in the gap values for Sn below a particle size of 20 nm. For these small sizes, gap values differ even more than 100% for similar sized Sn particles and enhancements as large as 60% with respect to the Sn bulk gap are found. In both systems however, superconductivity is destroyed below a critical particle size which is consistent with the Anderson criterion,[2] according to which superconductivity should be completely destroyed for particle sizes where the mean level spacing becomes equal to the bulk superconducting energy gap (see supplementary information)[20]. It is also worth noting that the average gap for the large Sn nanoparticles (20-30 nm) shows an increase of 20% from the bulk value (See supplementary information).

From the two parameters characterizing the superconducting state of our nanoparticles, $\Delta$ and $\Gamma$, only $\Gamma$ evolves in a similar way as a function of particle size both for Pb and Sn (Fig. 2**g**). In both systems, we observe an increase in $\Gamma$ with reduction in particle size. Interestingly, it seems that superconductivity is limited to sizes where $\Gamma < \Delta_{bulk.}$. At smaller sizes superconductivity is completely suppressed in both systems. This indicates that $\Gamma$ may have a particular significance in our measurements. To understand the behaviour of $\Gamma$ with particle size we invoke the role of quantum fluctuations in small particles. It is known from both theoretical calculations and experiments that there should be an increase in the quantum fluctuations in confined geometries[21,22,23] as observed by Bezryadin *et al* in their experiments on nanowires[8]. Similarly, since in a zero dimensional superconductor the number of electrons taking part in superconductivity decreases, we expect an increase in the uncertainty in the phase of the superconducting order parameter[18,19](within a single particle, there will be a decrease in the long range phase coherence). The increased fluctuations in the superconducting order parameter are expected to increase $\Gamma$ (as fluctuations act as a pair breaking effect). Therefore, we associate $\Gamma$ with the energy scale related with quantum fluctuations. Our results indicate that in zero dimensional systems the presence of quantum fluctuations of the phase (where $\Gamma > \Delta_{bulk}$) set the limit to superconductivity and this corresponds to the size consistent with the Anderson criterion[23].

We focus now on the main result of this work, reflected in the variation of $\Delta$ with particle size in Sn nanoparticles, and the observed striking difference with Pb. In order to interpret the experimental results we carry out a theoretical study of finite size corrections in the BCS formalism in line with references 4, 5 and 6. We will primarily focus only on the finite size corrections to the BCS gap equation since the corrections to the BCS mean field approximation[5] leads to a monotonic decrease in the gap[24] and are not responsible for the observed oscillations in Sn nanoparticles. For the correction to the BCS gap equation, two types of corrections are identified, smooth and fluctuating[4,5]. The former depends on the surface and volume of the grain and always enhances the gap with respect to the bulk. Since this contribution decreases monotonically with the system size it is not relevant in the description of the experimental fluctuations of $\Delta$. In order to explain the observed fluctuations of gap in Sn, we start with the self consistent equation for the BCS order parameter[5,6],

$$\Delta(\varepsilon) = \int_{-E_D}^{E_D} \frac{\lambda \Delta(\varepsilon') I(\varepsilon, \varepsilon')}{2\sqrt{\varepsilon'^2 + \Delta(\varepsilon')^2}} \frac{\nu(\varepsilon)}{\nu(0)} d\varepsilon' \qquad (3)$$

where $I(\varepsilon,\varepsilon') = V\int_0^L \Psi_\varepsilon^2(r)\Psi_{\varepsilon'}^2(r)dr$, $E_D$ is the Debye energy, L is a typical length of the grain, $\nu(0)$ is the spectral density at the Fermi level, $\lambda$ is the dimensionless coupling constant, $\nu(\varepsilon) = \sum_i g_i \delta(\varepsilon - \varepsilon_i)$ where $\varepsilon_i$ are the eigenvalues, with degeneracy $g_i$, and $\psi_\varepsilon(r)$ are the eigenfunctions with energy $\varepsilon$ of a free particle confined inside the grain. For Sn, a weak coupling superconductor a simple BCS formalism is capable of providing a good quantitative description of superconductivity. Eqn. 3 can be further simplified by noting[4,5,6] that for $k_F L \gg 1$ gap oscillations are controlled only by $\nu(\varepsilon)$. In our experiment (where *L* ranges between 2-60 nm) we are always in this limit as the Fermi wave vector $k_F = 16.4$ nm$^{-1}$ in Sn. As explained in the introduction, the gap oscillations arise from the discreteness of the level spectrum (see Fig. 1) which is reflected in the expression of the spectral density $\nu(\varepsilon)$ and hence Eqn. 3 leads to an oscillatory variation of gap with particle size. It can also be seen from the expression of $\nu(\varepsilon)$ that the presence of degenracies ($g_i > 1$) will enhance the gap fluctuations. Large $g_i$ is typical of grains with symmetry axes in which the energy levels are degenerate in a quantum number. A typical example is the sphere with three axes of symmetry. In this case each level in the energy spectrum with an angular momentum quantum number *l* is 2l+1 times degenerate.

We next proceed to solve Eqn. 3 numerically. Since we are only interested in fluctuations, for simplicity in the calculations we will set $I(\varepsilon,\varepsilon') = 1$. An important parameter in Eqn. 3 is $\lambda$ which implies an effective coupling constant (electron phonon coupling minus the coulomb repulsion) providing strictly within the BCS formalism a quantitative description of the superconductor. A natural choice is $\lambda = 0.25$ (for Sn) as this leads to the bulk gap and the coherence length consistent with the experimental values of these observables. The magnitude of the fluctuations will strongly depend on the shape of the grain as expected from the theory of shell effects[5]. From the experimental topographic images of the nanoparticles we can infer that the shape is very close to being a hemisphere (It cannot be said with certainty since the diameter of the particle is convoluted with the tip radius). However a statistical analysis of the nanoparticle images reveals that the deviations from an ideal hemispherical shape should not be larger than 15%. Hence for calculations, we model the shape of the nanoparticles as being a spherical cap with h/R > 0.85. We solve Eqn. 3 numerically after computing the $\varepsilon_i$ for a given ratio of h/R. In the hemispherical case, h/R = 1, the eigenvalues are simply the roots of a Bessel function. For other ratios, we use a method based on a perturbative expansion around the hemispherical geometry which is only valid for 1 - h/R << 1 (similar to the treatment in Ref. 25). The parameters used to describe the Sn nanoparticles are the height, $h$, measured by the STM, $k_F$ = 16.4 nm$^{-1}$, $E_F$ = 10.2 eV, $E_D$ = 9.5 meV and the coupling constant $\lambda$ = 0.25. We plot the calculated normalized gap as obtained from Eqn. 3 as a function of $h$ (calculations done down till h = 10 nm to safely remain within the validity of the BCS formalism) (solid lines in Fig. 3**a**) and superimpose the experimental results of Sn nanoparticles from Fig. 2**f** (shown by solid symbols in Fig. 3**a**). Here the data is normalized with respect to the average gap value obtained experimentally[26]. For h/R ranging between 0.9 to 0.95 (see supplementary information) we obtain a reasonably good quantitative matching with the theoretical results, indicating that finite size corrections can satisfactorily explain the results of Sn nanoparticles.

The natural question which follows is why such oscillations in Δ are not observed for Pb nanoparticles (solid triangles in Fig. 2**f**) (Note that oscillations in the gap have been observed in 2D Pb thin films below a critical thickness of 2 nm (thickness < Fermi wavelength) as a function of the number of layers in the film (Ref. 11-13). This phenomenon originates from the quantum confinement in the *z* direction leading to an oscillatory behavior of the density of states at Fermi level (with infinite degeneracy of the levels along the other two directions) and is independent of the superconducting coherence length). We recall that fluctuations in 0D systems have its origin in the discreteness of the spectrum and any mechanism that induces level broadening will suppress these oscillations. The superconducting coherence length (ξ) of Pb (~80 nm)[27] being much shorter than that of Sn (~240 nm)[27] will introduce a level broadening (broadening ∝ $v_F/\xi$). Moreover, since interactions are much stronger in Pb, the lifetime of the quasiparticles is shorter and an additional level broadening is expected. In Fig. 3**b** we plot the average oscillations obtained from both experiments and theory as a function of particle height for Pb and Sn nanoparticles. These average oscillations are the standard deviation of the gap from the average value[26]. We observe a good matching between theory and experiments. We would like to point out that for Pb the BCS description is an oversimplified model and one needs to solve the Eliashberg equations[28] to obtain the correct average gap values. However, to compute the oscillations in the gap and to check the suppression of the shell effects, BCS gives a reasonably good description for the strong coupling Pb (Fig. 3**b**).

Our results indicate that for any classical BCS superconductor with large quantum coherence lengths it is possible to enhance the superconducting energy gap by large factors (~60%) by tuning only the particle size. This may prove to be very useful in case of fullerides or hexaborides which are known to show a relatively high $T_c$ in the bulk.

**Methods:**

The experiments were performed in an ultra high vacuum (UHV) system (base pressure < $5 \times 10^{-11}$ Torr) combined with a home-built low temperature STM. The base temperature was 0.9 K. Differential conductance (dI/dV) spectra were measured with a tungsten (W) tip with open feedback loop using the lock-in technique with a 50 µV voltage modulation. A stabilization current of 0.1 nA and an initial sample voltage of 8.0 mV was used to measure all the tunneling spectra. The Pb and Sn nanoparticles of 1-35 nm height were grown *in situ* on top of a BN/Rh(111) surface by means of buffer layer assisted growth[29] (see Fig. 2**b**) where the BN having a band gap of ~6 eV acts as a decoupling layer. The topographic images of the nanoparticles on the surface were taken with the STM. The 3D plots were obtained by using the WSxM software.[30]

**References:**

[1] Von Delft, J. Superconductivity in ultrasmall metallic grains. *Annalen der Physik* **10**, 219-276 (2001).

[2] Anderson, P. W. Theory of dirty superconductors. *J. of Phys. and Chem. of Solids* **11**, 26-30 (1959).

[3] Kresin, V. Z. & Ovchinnikov, Y. N. Shell structure and strengthening of superconducting pair correlation in nanoclusters. *Phys. Rev. B* **74**, 024514-11 (2006).

[4] Heiselberg, H. Pairing of fermions in atomic traps and nuclei. *Phys. Rev. A* **68**, 053616-10 (2003).

[5] Garcia-Garcia, A. M., Urbina, J. D., Yuzbashyan, E. A., Richter, K. & Altshuler, B. L. Bardeen-Cooper-Schrieffer Theory of Finite-Size Superconducting Metallic Grains. *Phys. Rev. Lett.* **100**, 187001-4 (2008).

[6] Olofsson, H., Aberg, S. & Leboeuf, P. Semiclassical Theory of Bardeen-Cooper-Schieffer Pairing-Gap fluctuations. *Phys. Rev. Lett*, **100**, 037005-4 (2008).

[7] Ralph, D. C., Black, C. T. & Tinkham, M. Spectroscopic Measurements of Discrete Electronic States in Single Metal Particles. *Phys. Rev. Lett.* **74**, 3241-3244 (1995).

[8] Bezryadin, A., Lau, C. N. & Tinkham, M. Quantum suppression of superconductivity in ultrathin nanowires. *Nature* **404**, 971-974 (2000).

[9] Shanenko, A. A., Croitoru, M. D., Zgirski, M., Peeters, F. M. & Arutyunov, K. Size-dependent enhancement of superconductivity in Al and Sn nanowires: Shape-resonance effect. *Phys. Rev. B* **74**, 052502-4 (2006).

[10] Qin, S., Kim, J., Niu, Q. & Shih, C.-K. Superconductivity at the Two-Dimensional Limit. *Science* **324**, 1314-1317 (2009).

[11] Guo, Y. *et al.* Superconductivity Modulated by Quantum Size Effects. *Science* **306**, 1915-1917 (2004).

[12] Zhang, F-Y. *et al*, Band Structure and Oscillatory Electron-Phonon Coupling of Pb Thin Films Determined by Atomic-Layer-Resolved Quantum-Well States. *Phys. Rev. Lett.* **95**, 096802-4 (2005).

[13] Shanenko, A. A., Croitoru, M. D. & Peeters, F. M. Quantum size effects on $T_c$ of superconducting nanofilms. *Europhys. Lett.* **76**, 498-504 (2006).

[14] Brun, C. *et al*,. Reduction of the Superconducting Gap of Ultrathin Pb Islands Grown on Si(111). *Phys. Rev. Lett.* **102**, 207002-4 (2009).

[15] Bose, S., Raychaudhuri, P., Banerjee, R., Vasa, P. & Ayyub, P. Mechanism of the Size Dependence of the Superconducting Transition of Nanostructured Nb. *Phys. Rev. Lett.* **95**, 147003-4 (2005).

[16] Li, W. H. *et al.* Coexistence of ferromagnetism and superconductivity in Sn nanoparticles. *Phys. Rev. B* **77**, 094508-7 (2008).

[17] Brihuega, I. *et al*, Quantum and critical fluctuations in the superconductivity of single, isolated Pb nanoparticles. arXiv:0904.0354v1 (2009).

[18] Tinkham, M. *Introduction of Superconductivity,* 2nd edition (McGraw Hill, Singapore, 1996).

[19] Dynes, R. C., Narayanamurti, V. & Garno, J. P. Direct Measurement of Quasiparticle-Lifetime Broadening in a Strong-Coupled Superconductor. *Phys. Rev. Lett.* **41**, 1509-1512 (1978). This equation is phenomenological though it works in many different systems.

[20] In recent years this criteria has been substantially refined where it is accepted that superconductivity is destroyed at sizes depending on the parity of the grain and can be lower than that predicted by the Andersons criterion. See, Delft, V-J., Zaikin, A.D., Golubev, A. S. & Tichy, W. Parity-affected superconductivity in ultrasmall metallic grains, *Phys. Rev. Lett.* **77**, 3189-3192 (1996). We think that we are not sensitive to the parity of the particle.

[21] Dynes, R. C., Garno, J. P., Hertel, G. B. & Orlando, T. P. Tunneling Study of Superconductivity near the Metal-Insulator Transition. *Phys. Rev. Lett.* **53**, 2437-2440 (1984).

[22] Skocpol, W. J. & Tinkham, M. Fluctuations near superconducting phase transitions. *Rep. on Prog. in Phys.*, 1049-1097 (1975).

[23] Bennemann, K. H. & Ketterson, J. B. *Superconductivity:Conventional and Unconventional Superconductors*, *Volume 1* (Springer, 2008).

[24] The leading correction to the BCS prediction is given by $\Delta = \Delta_{BCS} - \delta/2$ where $\Delta_{BCS}$ is the bulk gap and $\delta$ is the discrete energy level spacing of the nanoparticle. This predicts a decrease in the average gap with reduction in particle size and can qualitatively explain the monotonic dependence of the average gap in both Pb and Sn nanoparticles.

[25] Rodríguez, A. H., Trallero-Giner, C., Ulloa, S. E. & Marín-Antuña, J. Electronic states in a quantum lens, *Phys. Rev. B* **63**, 125319-9 (2001).

[26] We divided the particle size in small bins of 2 nm wide and the average of the superconducting energy gap in each bin was determined.

[27] Kittel, C. *Introduction to Solid State Physics*, 8th edition (Wiley, 2004).

[28] Bergmann, G. & Rainer, D. The sensitivity of the transition temperature to changes in $\alpha^2 F(\omega)$. *Z. Phys*, **263**, 59-68 (1973).

[29] Berner, S. *et al.* Boron Nitride Nanomesh: Functionality from a Corrugated Monolayer. *Angew. Chem. Int. Ed.* **46**, 5115-5119 (2007).

[30] Horcas, I. *et al.* WSXM: A software for scanning probe microscopy and a tool for nanotechnology. *Rev. of Sci. Inst.* **78**, 013705-8 (2007).

**Acknowledgements**

We will like to thank Prof. K. Richter and Dr. M. Ternes for critically reading the manuscript. S.B. would like to thank the Alexander von Humboldt foundation and I. B. the Marie Curie action for support. A. M. G. G acknowledges financial support from the Spanish DGI through project No. FIS2007-62238.

**Author contributions**

S.B., I.B. and K.K. designed the research. S.B. and I.B. performed the experiments supervised by K.K.. A.M.G.G and J.D.U. provided the theoretical support. M.M.U and C.H.M helped in the experiments. S.B. and I.B. analyzed the data. The paper was written by S.B. and A.M.G.G. All authors contributed to the scientific discussion and revised the manuscript.

**Additional information**

The authors declare no competing financial interests. Correspondence and requests for materials should be addressed to S.B, I. B and A. M. G. G.

**Figure Captions:**

**Figure 1 | Schematic of Shell effects**

Schematic explaining the physical origin of shell effects in small particles which leads to an oscillation in the gap value with particle size. The left panel shows the energy band diagram of a small particle with a height $h$ where the discretization of the energy levels is arising from quantum confinement. Also for a particle with definite axes of symmetry, each level has further degeneracies and each degenerate level in a small particle is referred to as the 'SHELL'. Now, in superconductivity only the levels within the pairing region (Debye window) about the Fermi level, $E_F$ takes part in pairing and consequently superconductivity. We show the expansion of this pairing region for three particles with heights $h_1$, $h_2$ and $h_3$ which are very close to each other (so that the mean level spacing is quite similar). The number of levels in this pairing window fluctuates depending on the position of the Fermi level in the three particles which leads to the fluctuation in the gap (Shell effects).

**Figure 2 | Experimental configuration and low temperature superconducting properties of single, isolated Pb and Sn nanoparticles: Observation of Shell effects**

**a**, 3D representation of the experimental set-up. Superconducting nanoparticles deposited on a BN/Rh(111) substrate vary in height between 1-35 nm and are probed individually with the help of the STM tip.

**b**, 125 X 90 nm$^2$ 3D STM image showing the Sn nanoparticles of varying sizes deposited on the BN/Rh (111) substrate. The scale bar is shown in the left. The image is taken at a sample bias voltage of 1 V with a tunneling current of $I_t$ = 0.1 nA. This is a representative of topographic images of the superconducting Pb and Sn nanoparticles on the substrate.

**c-e**, Normalized conductance spectra (dI/dV *vs* V, normalized at a bias voltage of 5 mV). The circles are the raw experimental data and the solid lines are the theoretical fits using Eqns. 1 and 2 (see text). **c**, for Pb nanoparticles of different heights at T = 1.2 K. **d**, for two large Sn nanoparticles with heights 29.5 and 29.0 nm at T = 1.4 K. **e**, for two small Sn nanoparticles with heights 10.5 and 10.0 nm at T = 1.4 K.

**f,g** Comparison of the variation of superconducting energy gap (Δ) and broadening parameter (Γ) at low temperature (T = 1.2 K-1.4 K) for different Pb and Sn nanoparticles respectively as a function of particle height. The gap is normalized with respect to the bulk gaps. Observation of Shell effects in Sn nanoparticles. The solid lines in **f** are guides to the eye.

**Figure 3 | Comparison of experimental results with theoretical calculations obtained from finite size corrections to the BCS model.**

**a**, Variation of normalized gap with particle height. The solid symbols are obtained from the experimental data and the solid line is obtained from the theoretical calculations. The oscillations in the gap are explained on the basis of shell effects obtained from finite size corrections to the BCS model.

**b**, Variation in the average oscillations in the gap for Pb and Sn with particle height. The solid symbols are experimental data while the dashed lines are obtained from the theoretical calculations.

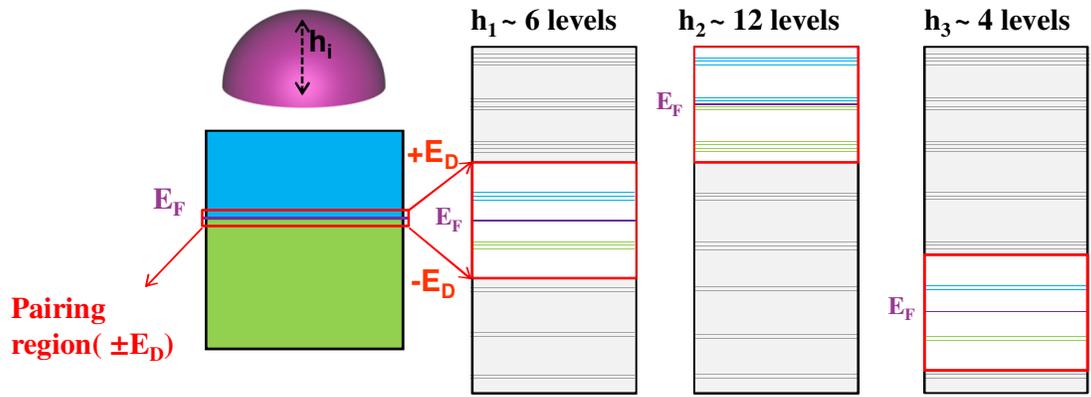

Figure 1

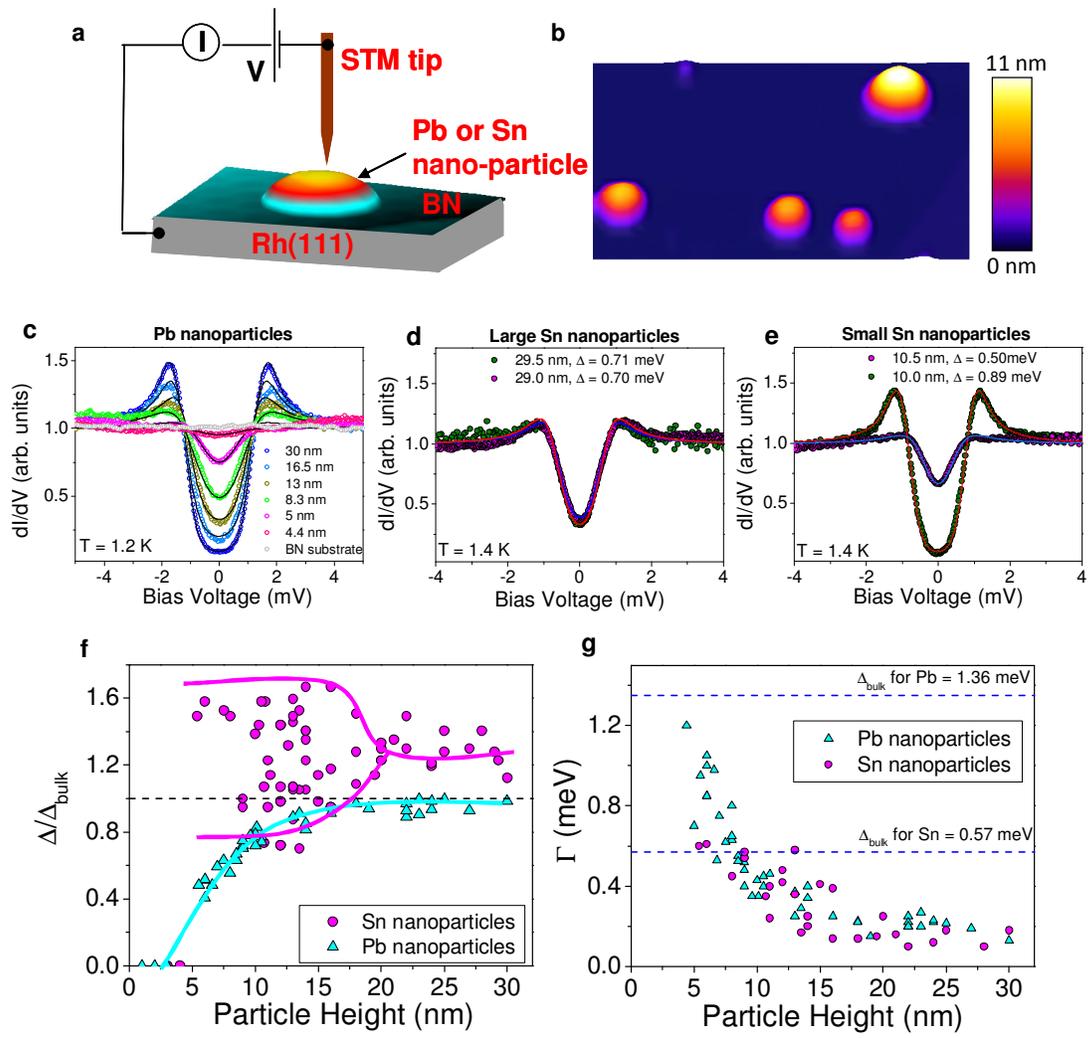

Figure 2

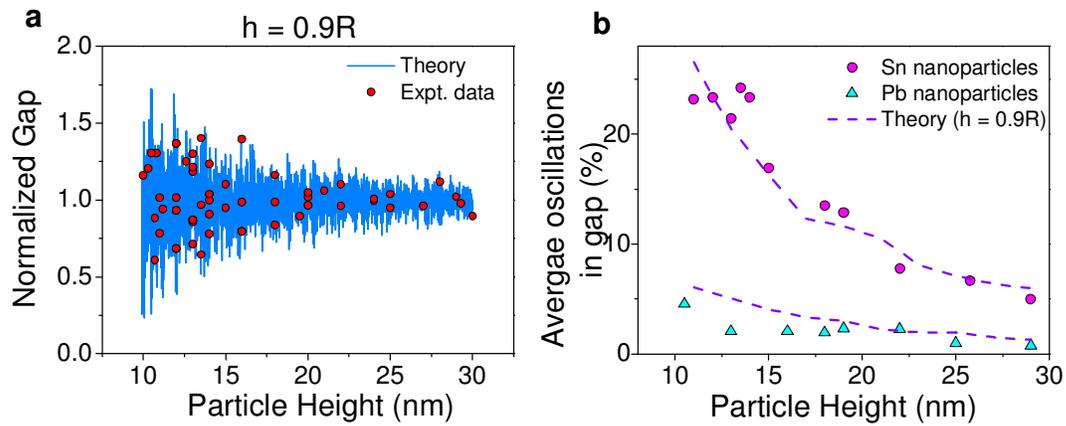

Figure 3